\documentclass{emulateapj}
\newcommand{\bm}[1]{\mbox{\boldmath $#1$}}
\begin{document}

\title{Formation Criteria and the Mass of Secondary Population III Stars}
\author{Hajime Susa\altaffilmark{1}}
\affil{Department of Physics, Konan University, Okamoto, Kobe, Japan}
\email{susa@konan-u.ac.jp}
\author{Masayuki Umemura\altaffilmark{2}}
\affil{Center for Computational Sciences, University of
  Tsukuba, Japan}
\email{umemura@ccs.tsukuba.ac.jp}
\author{Kenji Hasegawa\altaffilmark{2}}
\affil{Center for Computational Sciences, University of
  Tsukuba, Japan}
\email{hasegawa@ccs.tsukuba.ac.jp}

\begin{abstract}
We explore the formation of secondary Population III (Pop III) stars under 
radiation hydrodynamic (RHD) feedback by a preformed massive star.
To properly treat RHD feedback,
we perform three-dimensional RHD simulations incorporating
the radiative transfer of ionizing photons as well as H$_2$ 
dissociating photons from a preformed star.
A collapsing gas cloud is settled at a given distance 
from a 120$M_\odot$ Pop III star, and the evolution of the cloud 
is pursued including RHD feedback. 
We derive the threshold density depending on the distance, 
above which the cloud can keep collapsing owing to the shielding of 
H$_2$ dissociating radiation. 
We find that an H$_2$ shell formed ahead of an ionizing front works 
effectively to shield the H$_2$ dissociating radiation,
leading to the positive feedback for the secondary Pop III star formation. 
Also, near the threshold density, 
the envelope of gas cloud is stripped significantly by 
a shock associated with an ionizing front. 
By comparing the mass accretion timescale with
the Kelvin-Helmholtz timescale, we estimate the mass of secondary
Pop III stars. It turns out that the stripping by a shock can reduce
the mass of secondary Pop III stars down to $\approx 20M_\odot$.

\end{abstract}


\section{Introduction}
\label{intro}
The formation of Population III (hereafter Pop III)  
stars has been explored extensively
by many authors
\citep{Bromm99,Bromm02,NU99,NU01,Abel00,Abel02,
Yoshida06a,Yoshida06b,Oshea07},
and it is expected that the mass function of Pop III stars
is top-heavy with the peak of $100-1000M_\odot$.
But, the mass range of Pop III stars is not so clear. 
The enhanced H$_2$ cooling in pre-ionized gas may reduce
the mass \citep{SK87,SUNY98,OhH02}.
If the HD cooling is efficient in fossil HII regions,
the mass could be some $10M_\odot$
\citep[e.g.,][]{UI00, NU02,Nagakura05,JB06,GB06,Yoshida07}. 
Also, the variations of cosmological density fluctuations
may lead to the formation of less massive Pop III stars with 
some $10M_\odot$ \citep{Oshea07}.
However, the radiation hydrodynamic effect on the mass of secondary 
Pop III stars forming nearby a preformed massive star
has not hitherto explored in detail. 
From an observational point of view, 
the elemental abundance patterns of hyper-metal-poor stars 
well match the yields by supernova explosions with a 
progenitor mass of $\sim 25M_{\odot}$ \citep{Umeda03, Iwamoto05}. 
This implies that Pop III stars of some $10M_\odot$ are likely to have 
formed before the first metal enrichment. 

Metal-free first objects are expected to
collapse at $20 \lesssim z \lesssim 30$, 
forming a minihalo with the mass of $\approx 10^6 M_\odot$ and 
the extent of $\approx 100$pc
\citep{Tegmark97,NS99,Fuller00,Yoshida03}.
In the course of bottom-up structure formation, 
such minihaloes merge to form first galaxies at $z \gtrsim 10$,
having virial temperature $\gtrsim 10^4$K and mass $\gtrsim 10^8M_\odot$.
Even in the evolution of first galaxies, Pop III stars can play a significant role, 
since an appreciable number of stars can form from metal-free 
component in interstellar gas \citep{Tornatore07,Johnson08}.
Also, those Pop III stars are likely to be responsible 
for the reionization of the universe 
\citep{Cen03,Cia03,Wyi04,Somer03,Sokasian04,Murakami05},
and the metal enrichment of intergalactic medium 
\citep{NU01,Scan02,Ricotti04}
through supernova (SN) explosions 
\citep{Mori02,Bromm03,KY05,Greif07}.
Hence, the mass range of Pop III stars affects 
the galaxy formation and the evolution of intergalactic matter.

If first stars are very massive, they emit intensive ultraviolet (UV) 
radiation, which significantly influences the subsequent
star formation in minihaloes or first galaxies 
\citep{Haiman97,ON99,Haiman00,GB01,Machacek01}.
In a Pop III minihalo, primordial fluctuations generate 
density peaks owing to gravitational instability induced by
H$_2$ cooling. 
A highest peak collapses earlier to form a first Pop III star
as shown by \citet{Abel02}.
Subsequently, lower peaks collapse to form cloud cores 
with the density of $10^{2-4}$cm$^{-3}$ 
and the temperature of $200-500$K \citep{Bromm02}.
Hence, later collapsing cores could be affected by 
the radiative feedback by the first star. 
In first galaxies, the interstellar medium can be
pre-ionized by shock, since the virial temperature is higher than $10^4$K.
The pre-ionization can promote rapid H$_2$ formation 
\citep{SK87, KS92, SUNY98, OhH02}. As a result,
the temperature of interstellar gas in first galaxies can be lowered 
down to $100$K through efficient H$_2$ cooling
\citep{Johnson08}.
Low temperature collapsing clouds in first galaxies could be 
subject to the UV feedback by Pop III stars.

If a first star is distant by more than 1pc,
dense clouds are readily self-shielded from the ultraviolet (UV) radiation
\citep{TU98, Kitayama01,SU04a,SU04b,Kitayama04,Dijkstra04,Alvarez06}. 
Thus, the photoevaporation
by UV heating is unlikely to work devastatingly. 
However, Lyman-Werner (LW) band radiation (11.18-13.6eV) 
photodissociates H$_2$ molecules, and therefore
can preclude a cloud core from collapsing.
Hence, the photodissociation of H$_2$ may lead to momentous 
negative feedback 
\citep{Haiman97,ON99,Haiman00,GB01,Machacek01}.
Recently, \citet{Susa07} performed 3D radiation hydrodynamic
simulations on the photodissociation feedback, and derived 
the criteria of the feedback. 

However, ionizing radiation ($\geq$13.6eV) 
is related to H$_2$ formation in an intricate fashion. 
An ionization front (I-front) driven by ionizing radiation
propagates in a collapsing core. 
The enhanced fraction of electrons catalyzes rapid H$_2$ formation 
\citep{SK87, KS92, SUNY98, OhH02}. In particular,
the mild ionization ahead of the I-front can generate an H$_2$ shell,
which potentially shields H$_2$ dissociating photons \citep{Ricotti01}.
This mechanism is likely to work positively to form Pop III stars. 
When UV irradiates a dense core, the I-front changes from R-type 
on the surface to D-type inside the core. The transition occurs
via an intermediate type (M-type), which is accompanied with
the generation of shock \citep{Kahn54}. The shock can affect
significantly the collapse of the core. This is a totally radiation
hydrodynamic (RHD) process. Such radiation hydrodynamic feedback
has been investigated by 1D spherical RHD simulations 
\citep{AS07}, 2D cylindrical RHD simulations \citep{Whalen08},
and 3D RHD simulations \citep{SU06}. 
The results by 2D and 3D simulations are in good agreement with each other.
However, quantitative feedback effects, including the compression
and stripping by a shock, have not been explored yet over a wide
parameter range in 3D simulations. 

In this paper, we attempt to investigate quantitatively 
the RHD feedback on a collapsing cloud by a preformed first star. 
For the purpose, we solve the three-dimensional RHD 
in a wide range of parameter space. In the simulations, 
the radiative transfer of ionizing photons
as well as H$_2$ dissociating photons is self-consistently coupled
with SPH hydrodynamics. 
In \S 2, the numerical method is described. The setup of simulations is
presented in \S 3, and numerical results are given in \S 4.
In \S 5, the formation criteria of secondary Pop III stars
are numerically derived taking RHD feedback into account.
In \S 6, the mass of secondary Pop III stars is estimated. 
\S 7 is devoted to the conclusions and discussion.

\section{Numerical Method}
We perform numerical simulations with the radiation hydrodynamics code that 
we developed \citep{Susa06}. In this section, we briefly summarize the scheme. 
The code is designed to investigate the formation and evolution of the first-generation 
objects at $z \gtrsim 10$, where the radiative feedback from various sources plays 
important roles. The code can compute the fraction of the chemical species 
e$^-$, H$^+$, H, H$^-$, H$_2$, and H$_2^+$ 
through fully implicit time integration. 
It also can deal with multiple sources of ionizing radiation, as well as the radiation 
at the Lyman-Werner band.
Although there may be heavy elements in first galaxies
through Pop III supernova explosions therein,
the metallic cooling is not dominant as long as the metallicity is lower
than $10^{-2}Z_\odot$ \citep{SU00}. In this paper, we ignore the metallic cooling. 

The hydrodynamics is calculated by the smoothed particle hydrodynamics (SPH) method. 
We use the version of SPH by \citet{Umemura93}, 
with the modification according to \citet{SM93}. 
We adopt the particle resizing formalism by \citet{Thacker00}.

The non-equilibrium chemistry and radiative cooling for primordial gas are 
calculated with the code developed by \citet{SK00}, 
where the H$_2$ cooling and reaction rates are mostly taken from \citet{GP98}. 
As for the photoionization process, we employ the so-called on-the-spot 
approximation \citep{Spitzer78}, where it is assumed that recombination photons 
to ground states of hydrogen are absorbed promptly on the spot,
while recombination photons to excited states can escape the medium.
In the present version of our code, we do not take into account helium
ionization. For pure hydrogen gas with on-the-spot approximation, 
we do not need the frequency bins for the radiation field. All we need is
the optical depth at Lyman limit, since we know the frequency dependence
of the optical depth. We can obtain the ``precise'' photoionization rate or
photoheating rate as functions of the optical depth at Lyman
limit before we start simulations \citep{Susa06}.  We also remark that
in case we include helium, we need more frequencies\citep{NUS01,Susa06}.

The optical depth is integrated using the neighbor lists of SPH particles. 
It is similar to the code described in \citet{SU04a}, but now we can deal 
with multiple point sources. In our new scheme, we do not create as many grid points 
on the light ray as the previous code \citep{SU04a} does. 
Instead, we just create one grid point per SPH particle located in its neighbor. 
We find the ``upstream'' particle for each SPH particle on its line of sight 
to the source. Then the optical depth from the source to the SPH particle is 
obtained by summing up the optical depth at the ``upstream'' particle and 
the differential optical depth between the two particles.

The opacity against LW-band flux ($F_{\rm LW}$)  is 
calculated with the self-shielding function by \citet{DB96}, which 
is given by
\begin{equation}
F_{\rm LW} = F_{\rm LW,0} f_{\rm sh}
\left( N_{\rm H_2,14 } \right) \label{LW}
\end{equation}
where $ F_{\rm LW,0}$ is the incident flux, 
$ N_{\rm H_2,14}= N_{\rm H_2}/10^{14} {\rm cm^{-2}}$
is the normalized H$_2$ column density, and
\begin{equation}
f_{\rm sh}(x) = \left\{
\begin{array}{cc}
1,~~~~~~~~~~~~~~x \le 1 &\\
x^{-3/4},~~~~~~~~~x > 1 &
\end{array}
\right.
\end{equation}
The column density of H$_2$ is evaluated by 
the method described above. We assume the same absorption for 
the photons within the LW-band between 11.26 and 13.6 eV.
The use of the self-shielding function ignores the effects of Doppler
shifts of LW lines. If the line is shifted more than thermal broadening
width, the opacity
at LW band is greatly reduced. In the present calculations, the
collapsing gas cloud converges to Larson-Penston type similarity
solutions. 
Thus, the infall velocity in the envelope of cloud is as large as
the sound velocity. Moreover, the velocity around the core is much smaller
than sound velocity. Therefore, we do not include this effect in the
present simulations.

The code is already parallelized with the MPI library. 
The computational domain is divided by the so-called orthogonal recursive 
bisection method \citep{Dubinski96}. 
The parallelization method for the radiation transfer part is similar to 
that of the multiple wave front method developed by \citet{NUS01} and 
\citet{Heinemann06}, but it is changed to fit the SPH scheme. 
The code is able to handle self-gravity with a Barnes-Hut tree 
\citep{BH86}, which is also parallelized. The code is tested for various 
standard problems \citep{Susa06}. We also take part in the code comparison 
project with other radiation hydrodynamics codes \citep{Iliev06,Iliev09}, 
and we find reasonable agreements with each other.

The present simulations are mainly carried out with 
a novel hybrid computer system in University of Tsukuba, 
called {\it FIRST} simulator, which has been designed to simulate 
multi-component self-gravitating radiation hydrodynamic systems
with high accuracy \citep{Umemura07}. 
The {\it FIRST} simulator is composed of 256 nodes with dual Xeon processors, 
and each node possesses a Blade-GRAPE board, 
that is, the accelerator of gravity calculations. 
The peak performance is 36.1 Tflops.

\section{Setup of Simulations}
\label{simulation}

We use $N=524,288(=2^{19}$) SPH particles to 
perform radiation hydrodynamic simulations.
We suppose a primordial gas cloud undergoing run-away collapse
nearby a preformed massive star.
Dark matter is neglected in the present calculations, because
we are interested in rather dense cores with
$n > 10^{2-4} {\rm cm^{-3}}$ that start to become baryon-dominated
(e.g. Abel, Bryan \& Norman 2002). 
The tidal field from other halos could have some effects on the dynamics of
collapsing gas, which is not taken into account in the present
calculation.
The chemical compositions are initially assumed to have 
the cosmological residual value \citep{GP98}. 
The mass of cloud is $M_{\rm b} = 8.3\times 10^4 M_\odot$ 
in baryonic component. The gas of the cloud is initially at rest and
have ``top-hat'' density distribution. We also assume initially
uniform temperature. 
Remark that the density distribution converges to ordinary $\rho \propto
r^{-2}$ profile as the collapse
proceeds (e.g. Abel, Bryan \& Norman 2002).
Also, we distribute a low-density 
($n \simeq 10^{-1}{\rm cm^{-3}}$) uniform gas 
around the collapsing cloud.
When the central density of cloud exceeds a certain value, 
$n_{\rm on}$, we ignite a 120$M_\odot$ Pop III star.
The luminosity and the effective temperature
of the source star are taken from \citet{Baraffe01}.
The ionizing photon number per unit time is
$\dot{N}_{\rm ion}=1.3\times 10^{50}{\rm s^{-1}}$, whereas the LW photon
number luminosity is $\dot{N}_{\rm LW}=1.4\times 10^{49}{\rm s^{-1}}$.
The source star is located at the distance $D$ from the center
of collapsing cloud. 

\begin{figure}
\begin{center}
\includegraphics[angle=0,width=7cm]{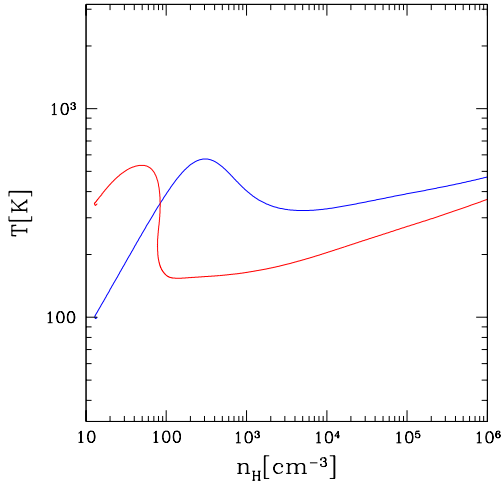}
\caption{Cloud collapse without radiative feedback.
The temperature at the center is shown against the central density 
for the initial temperature of 350K (red curve) and 100K (blue curve).
} \label{nofeedback}
\end{center}
\end{figure}

First, we see fundamental physical processes of radiative
feedback using high/low temperature core models. 
Then we derive the radiation hydrodynamic feedback criteria 
by changing $n_{\rm on}$ and $D$. 
The basic difference between two models
is the initial temperature of the clouds.
The high initial temperature results in the low temperature core  (LTC),
while the low initial temperature does in the high temperature core (HTC).
In Fig. \ref{nofeedback}, the cloud collapse in two cases
is shown by temperature versus density at the center.
In a LTC model, the relatively high initial temperature ($T_{\rm ini}=350$K)
allows the efficient formation of H$_2$ molecules and also leads to the
relatively slow gravitational contraction. In this model, the ratio
between gravitational energy to initial thermal energy $|W|/U\simeq 2$.
Consequently, the core temperature goes down quickly to
$\approx 150$K at the density $n_{\rm H}\approx 100{\rm cm}^{-3}$. 
Since the size of core
is approximately the Jeans length, a smaller core forms
in a LTC model. It is worth noting that the density distribution of
collapsing gas for $n_{\rm H}\ga 10^3 {\rm cm^{-3}}$ is close to the one
obtained in cosmological simulations including dark matter
gravity \citep{Oshea07,Susa07}. 
In an HTC model, the formation of H$_2$ molecules
is not so rapid owing to the lower initial temperature ($T_{\rm ini}=100$K).
Also, the initial low pressure ($|W|/U\simeq 3-4$) allows the
rapid gravitational contraction.
Hence, the temperature in a collapsing phase becomes higher, 
and eventually a larger core forms.
The difference of core size is significant for the radiative
feedback. 

The basic models investigated in this paper are shown in Table \ref{table1}.
LTC-ION and HTC-ION include both ionizing and LW-band radiation 
from a 120$M_\odot$ star respectively for LTC and HTC models, while
LTC-LW and HTC-LW include only LW-band radiation respectively for LTC and HTC models.
$T_{\rm c,min}$ is the minimum core temperature after the H$_2$ cooling instability.
In these models, $D=40$pc and $n_{\rm on}=10^3{\rm cm}^{-3}$ are assumed,
but $D$ and $n_{\rm on}$ are changed over a wide range 
to derive the feedback criteria. 

\begin{table}
   \caption{Basic Models}
\begin{center}
  \begin{tabular}{lccccc}
  \hline
  \hline
  Model & $T_{\rm c,min}$ [K] & Ionizing Flux & LW-band & $D$ [pc] & $n_{\rm on}$ [${\rm cm^{-3}}$] \\
   \hline 
  LTC-ION & 150 & Yes & Yes & 40 & $10^3$ \\
  LTC-LW & 150 & No & Yes & 40 & $10^3$ \\
  HTC-ION & 300 & Yes & Yes & 40 & $10^3$ \\
  HTC-LW & 300 & No & Yes & 40 & $10^3$ \\
   \hline \label{table1}
\end{tabular}
\end{center}
\end{table}

\section{Numerical Results}

\begin{figure*}[t]
\centering{
\includegraphics[angle=0,width=14cm]{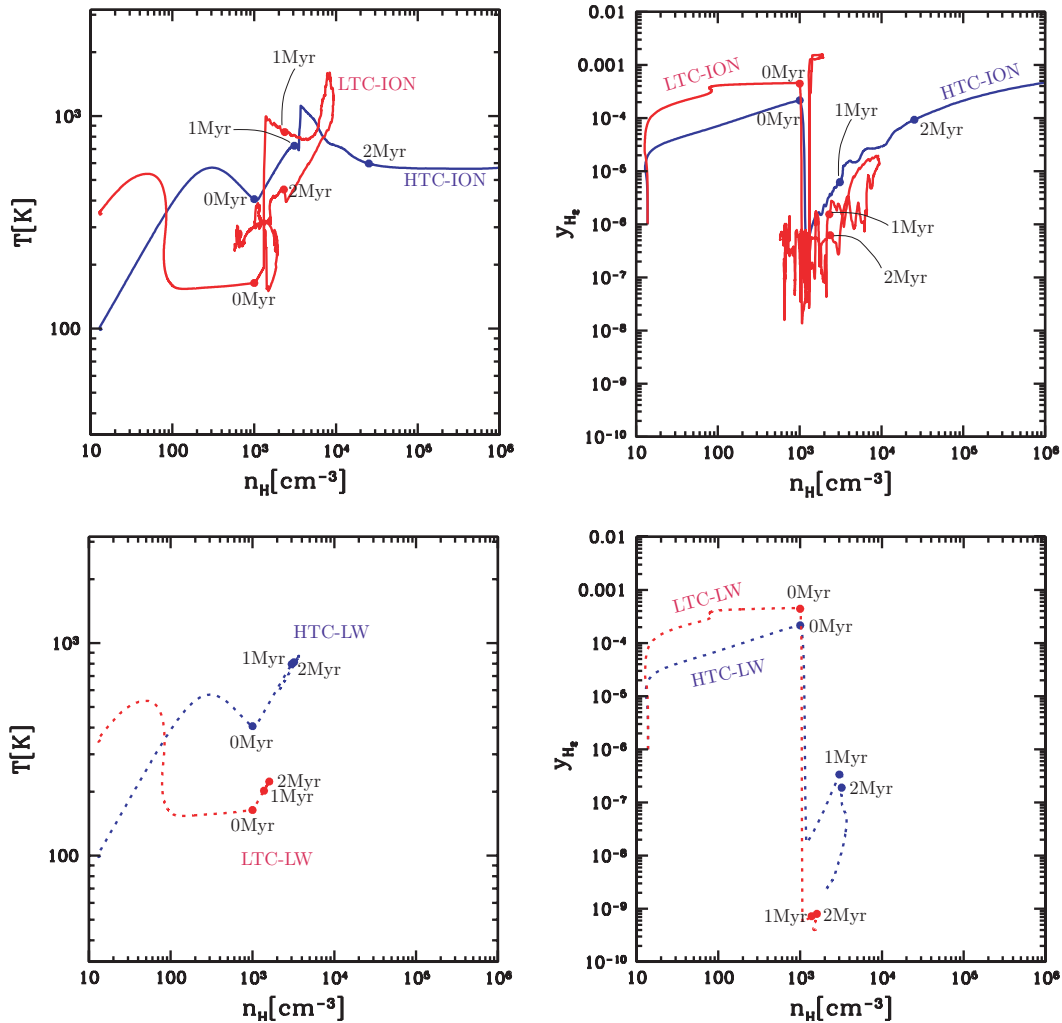}}
\caption{The cloud evolution with radiative feedback for the models
listed in Table \ref{table1}.
The time variations of temperature (left two panels) and H$_2$ molecule
abundance (right two) at the cores are shown against the peak density. 
A red solid line depicts LTC-ION model, a blue solid line does HTC-ION
 (upper two panels),
a red dotted line does LTC-LW, 
and a blue dotted line does HTC-LW (lower two panels).
On each curve three stages are indicated by dots, 
which correspond to 0yr, 1Myr, and 2Myrs after the ignition of a source star.
} \label{Evolution}
\end{figure*}

Here, we see the fundamental physical processes for
four basic models shown in Table \ref{table1}.

\subsection{Evolution of Cloud Cores}

The time evolutions of cloud cores for four models are shown 
in Fig. \ref{Evolution}. 
In upper two panels the temperature and H$_2$ molecule abundance
at the cores are shown against the peak density for LTC-ION and HTC-ION models, 
while in lower two panels they are shown for LTC-LW and HTC-LW models.

The cloud evolution is regulated by the formation of H$_2$ molecule, which
is an indispensable coolant for cloud collapse.
As seen in Fig. \ref{Evolution}, the H$_2$ abundance drops 
down to a level of $10^{-9}-10^{-6}$ immediately by UV radiation
after the ignition of a source star
(right panels, where the ignition time is marked as 0Myr). 
Then, owing to the inefficiency of H$_2$ cooling,
the gravitational contraction becomes adiabatic and temperature rises
until 1Myr (left panels). During the adiabatic contraction, 
H$_2$ molecules retrieve, since the H$_2$ formation rate is raised 
in adiabatically heated cores.

In HTC-ION model, the heating by a shock associated 
with an ionization front helps to increase the temperature. 
As a result, the H$_2$ formation is restored and 
H$_2$ fraction exceeds $\sim 10^{-4}$. 
Thus, the core can cool and keep collapsing.
In LTC-LW, and HTC-LW models, 
the H$_2$ abundance drops down to a level lower than $10^{-8}$ after UV
irradiation, and the temperature continues to increase owing to insufficient
cooling. Although H$_2$ formation is recovered, H$_2$ fractions are
not sufficient for the cloud to keep collapsing. 
Eventually, the cloud core bounces. 
In LTC-ION model, the physical processes are more complicated. 
The shock heating before 1Myr works importantly to restore H$_2$
formation. Resultantly, the H$_2$ abundance reaches to 
a level of $\sim 10^{-5}-10^{-4}$.
However, the second hit by a diffracted shock between 1Myr and 2Myr 
(the second peak in a red solid curve in the upper left panel 
in Fig.\ref{Evolution})
raises the temperature quickly, so that the core cannot be gravitationally
bounded and bounces eventually. 
Such intricate behavior of a shock is explained further in the following.

\subsection{Convergence check}
Ideally, it is appropriate to perform convergence check for all the
models, however, we check for HTC-ION model only,  because of the limited
computational time. 
We run three additional
simulations with particle number $N=2^{17},~ 2^{18},~2^{20}$, while the
canonical run uses $N=2^{19}$ particles.
\begin{figure*}[t]
\centering{
\includegraphics[angle=0,width=15cm]{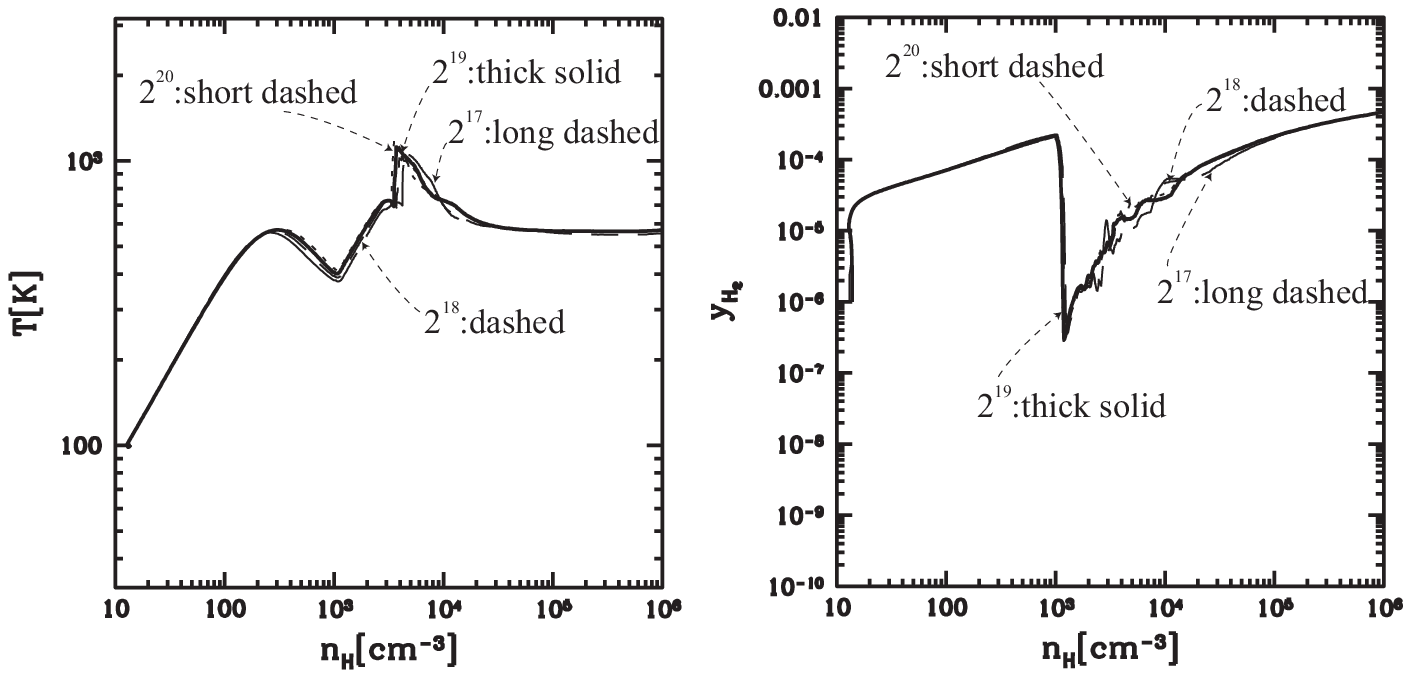}}
\caption{Convergence check on density-temperature plane (left panel) and
 density-H$_2$ abundance ($y_{\rm H_2}$) plane for HTC-ION model. In both of the panels,
 results from four runs are plotted. Short dashed: $N=2^{20}$, thick
 solid:$N=2^{19}$(canonical), dashed: $N=2^{18}$, long dashed:$N=2^{17}$ .
} \label{convergence}
\end{figure*}
Since the highest density peak collapses and we are now interested in the collapse criteria, 
we plot the evolution of the density
peak on density-temperature plane and density-H$_2$ abundance ($y_{\rm
H_2}$) plane 
for four different resolutions.
As shown in the Figure \ref{convergence}, all of the results agrees
fairly well, although the lowest resolution case (long dashed line)
shows some deviation from others. Thus, the the resolution of the present
simulations are enough to describe the collapse criteria. 

\subsection{Cloud Structure}

\begin{figure*}[t]
\centering{
\includegraphics[angle=0,width=15cm]{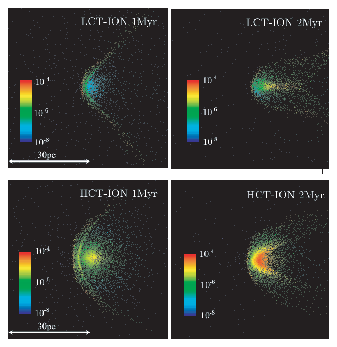}}
\caption{Hydrogen molecule distributions 
on a plane including the symmetry axis are shown at 1Myr and 2Myr 
after the UV irradiation for HTC-ION and LTC-ION models. 
The source star is located on the left
boundary of each panel. H$_2$ fractions are shown by colored dots
according to the color legend.
} \label{H2map}
\end{figure*}
\begin{figure*}[t]
\centering{
\includegraphics[angle=0,width=15cm]{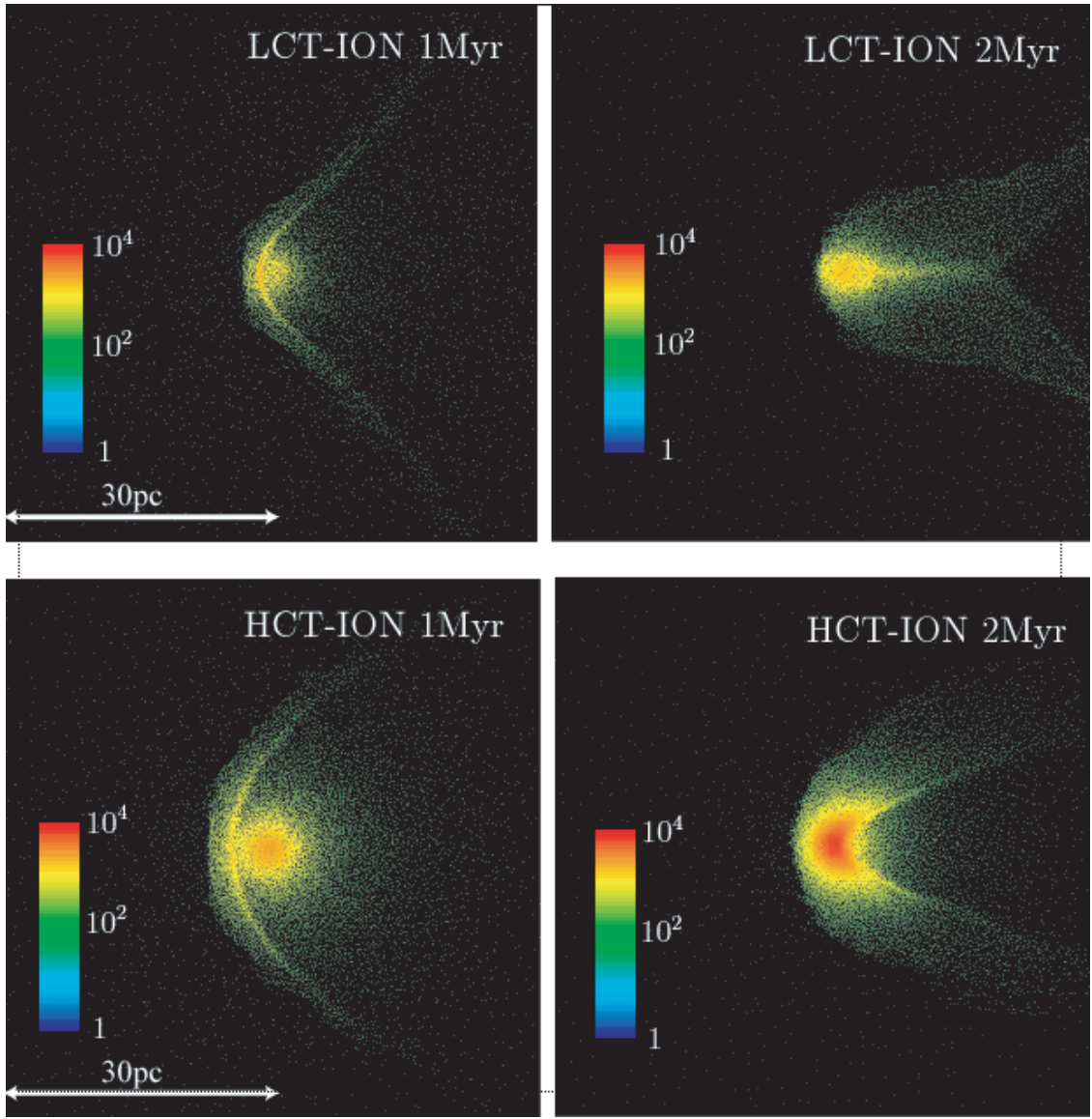}}
\caption{Same as Fig.\ref{H2map}, except that the colors of the dots
 represent the number density of hydrogen nucleous. 
 The color legend is depicted in units of cm$^{-3}$.
} \label{densitymap}
\end{figure*}
\begin{figure*}[t]
\centering{
\includegraphics[angle=0,width=15cm]{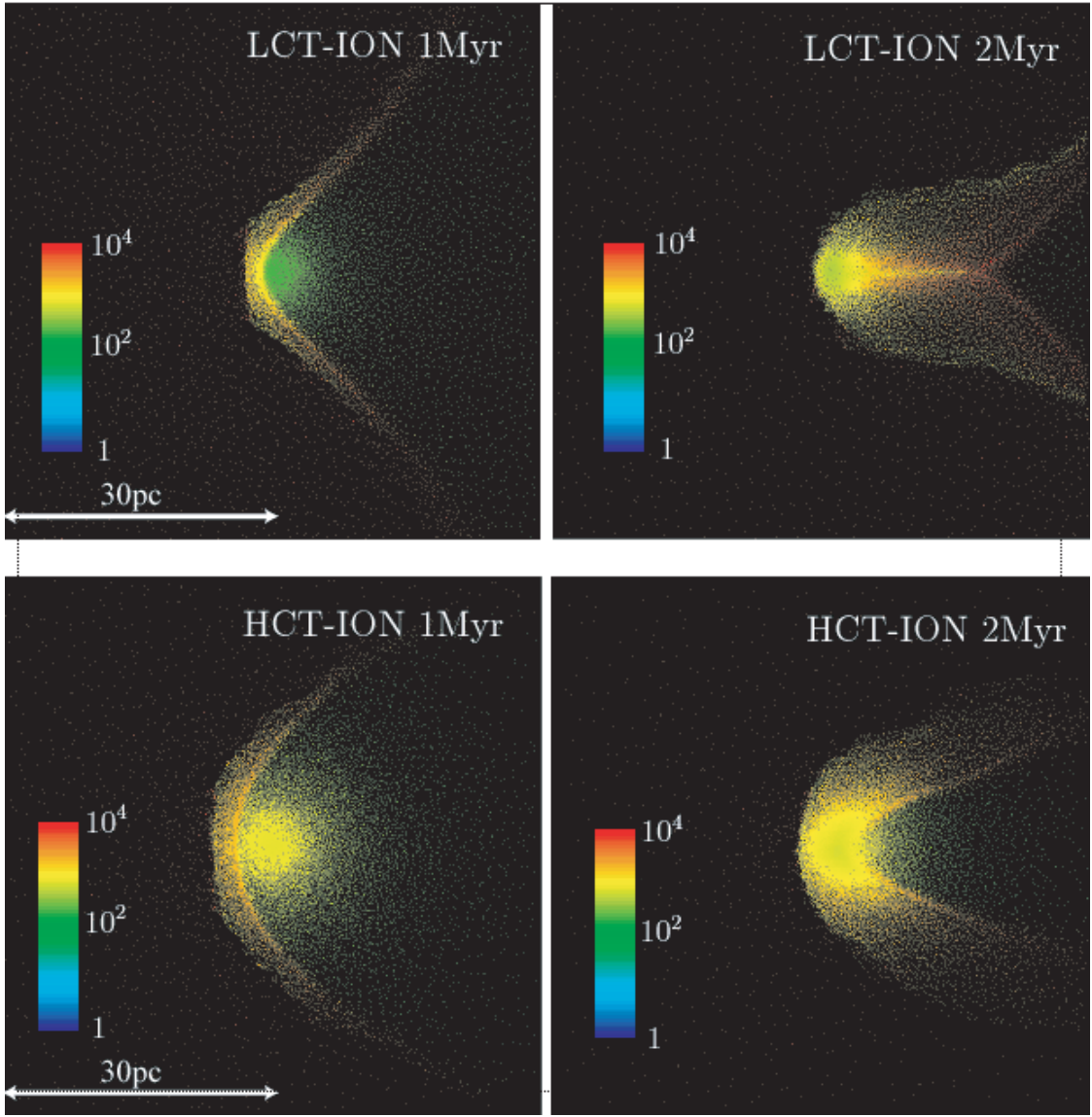}}
\caption{Same as Fig.\ref{H2map}, except that the colors of the dots
 represent the temperature of hydrogen nucleous. 
 The color legend is depicted in units of Kelvin. 
} \label{Tmap}
\end{figure*}
\begin{figure}[h]
\centering{
\includegraphics[angle=0,width=8cm]{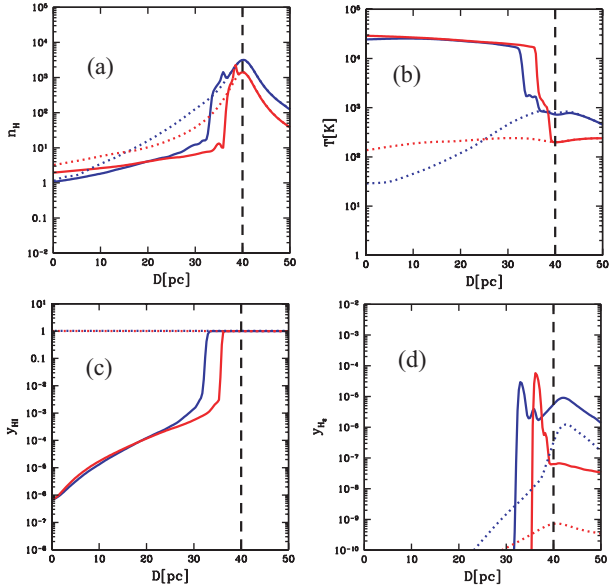}}
\caption{Various physical quantities at 1Myr after the UV irradiation 
along the symmetry axis are plotted.
Four panels represent (a) hydrogen number density $n_{\rm H} [{\rm
 cm^{-3}}]$ , (b) temperature $T$[K], 
(c) neutral hydrogen abundance $y_{\rm HI}$ , and (d) H$_2$ molecule
 abundance $y_{\rm H_2}$, respectively.
The horizontal axis shows the distance from the source star. 
A blue solid curve depicts HTC-ION model, a red solid curve does LTC-ION,
a blue dotted curve does HTC-LW, and a red dotted curve does LTC-LW.
A vertical dashed line in each panel shows the initial position of cloud center.
}\label{quantities}
\end{figure}

\begin{figure}
\begin{center}
\includegraphics[angle=0,width=7cm]{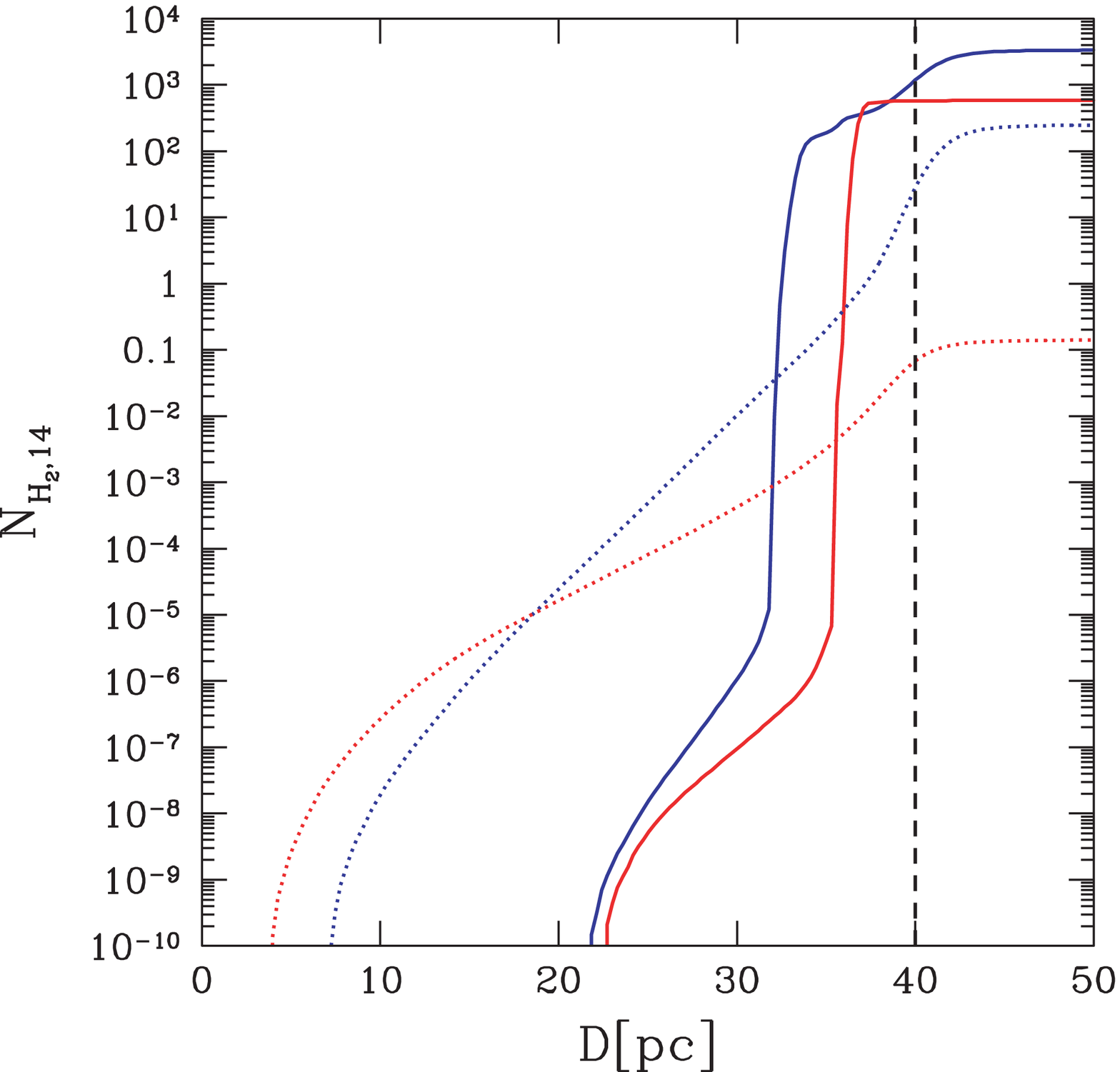}
\caption{
H$_2$ column density from the source star is shown
at 1Myr after the UV irradiation
in units of $10^{14}{\rm cm^{-2}}$.
The horizontal axis shows the distance from the source star. 
A blue solid curve depicts HTC-ION model, a red solid curve does LTC-ION,
a blue dotted curve does HTC-LW, and a red dotted curve does LTC-LW.
A vertical dashed line in each panel shows the position of cloud center.
}\label{nh2}
\end{center}
\end{figure}

The evolution shown above indicates that ionizing radiation works 
to promote H$_2$ formation, and therefore help the cloud to collapse.
In Fig. \ref{H2map}, 
the spatial distributions of H$_2$ molecules on the plane
including the symmetry axis are shown for LTC-ION and HTC-ION models.
Two snapshots are taken at the epoch of 1Myr and 2Myr after UV
irradiation. Figs.\ref{densitymap} and \ref{Tmap} show the gas number
density and temperature distributions.
Also, we plot various physical quantities along the symmetry axis
for all four models in Fig.\ref{quantities}. 

In LTC-ION model (upper two panels in Figs.\ref{H2map}-\ref{Tmap}), 
the core is self-shielded against ionizing photons 
and a bow shock forms preceding an ionization front.
The bow shock strips the envelope of a collapsing cloud.
H$_2$ molecules are produced efficiently ahead of an ionization
front owing to the enhanced catalytic effect of free electrons
(see also Fig.\ref{quantities}). 
As a result, an H$_2$ shell is built up and it shields effectively
the H$_2$ dissociating radiation from a source star. 
In Fig. \ref{nh2}, H$_2$ column density from
the source star is plotted at 1Myr after UV irradiation. 
In fact, $N_{\rm H_2}$ at the center of cloud
is several hundred times higher than the critical value 
$10^{14}{\rm cm^{-2}}$ for the shielding of LW-band radiation. 
Thus, H$_2$ dissociating radiation is
significantly absorbed before it reaches the cloud center.
However, as shown in the top-right panel in Figs.\ref{H2map}-\ref{Tmap}, 
a diffracted shock front hits the cloud core 
before undergoing the run-away collapse. 
Consequently, the core bounces owing to the thermal pressure 
enhanced by the shock.
It is also worth noting that we find ``an H$_2$-enriched tail'' behind the
core. This structure is formed by the collision of shock which is
diffracted around the dense core. The temperature of the shock heated
gas is high enough to feed electrons that catalyze the H$_2$
formation. 
(This structure could be interesting if it fragments into
small clumps and yields protostars, although
we do not find any sign of such a process 
in the present simulations. Regarding this issue, 
higher resolution simulations may be required.)

In HTC-ION model (lower two panels in Figs.\ref{H2map}-\ref{Tmap}), 
a bow shock forms, similar to LTC-ION model.
H$_2$ column density at an H$_2$ shell is comparable to
that in LTC-ION model. 
However, in this case, the core size is larger than LTC-ION model.
Thus, H$_2$ column density at the cloud center is roughly ten times
higher than that in LTC-ION model (see Fig. \ref{nh2}). 
Also, a diffracted shock is not converged into the cloud core before the
central density peak goes into run-away collapse phase. 
As a result, the cloud core can keep collapsing, 
although the cloud envelope is stripped by a bow shock. 

In LTC-LW model, a smaller core forms
owing to its lower temperature. 
Since the column density of H$_2$ molecules inside the cloud
is low, the self-shielding of H$_2$ dissociating radiation is very weak
(see Fig.\ref{quantities}).
As a result, the photodissociation of H$_2$ molecules is so intense
to reduce the H$_2$ abundance to $y_{\rm H2} \approx 10^{-8}$.
Consequently, the cloud core cannot cool and collapse. 

In HTC-LW model, although a large core forms 
owing to its higher temperature,
the dissociating radiation from a source star destructs
H$_2$ molecules to a level of H$_2$ abundance 
as $y_{\rm H2} \approx 10^{-6}$.
Hence, the H$_2$ cooling is not sufficient and 
the cloud core cannot collapse.  

To conclude, within these four models, only HTC-ION model
succeeds in collapsing. 
The other models result in bouncing after UV irradiation.
In other words, the positive feedback overwhelms the negative ones
in HTC-ION model, whereas the negative feedback hinders
cloud collapse in LTC-ION, LTC-LW, and HTC-LW models.
However, the fate of clouds depends on the distance $D$ 
from a source star and the core density $n_{\rm on}$ at UV irradiation. 
Hence, we investigate a wider parameter space 
to derive the formation criteria of secondary Pop III stars.

\section{Formation Criteria}

\begin{figure}
\begin{center}
\includegraphics[angle=0,width=7cm]{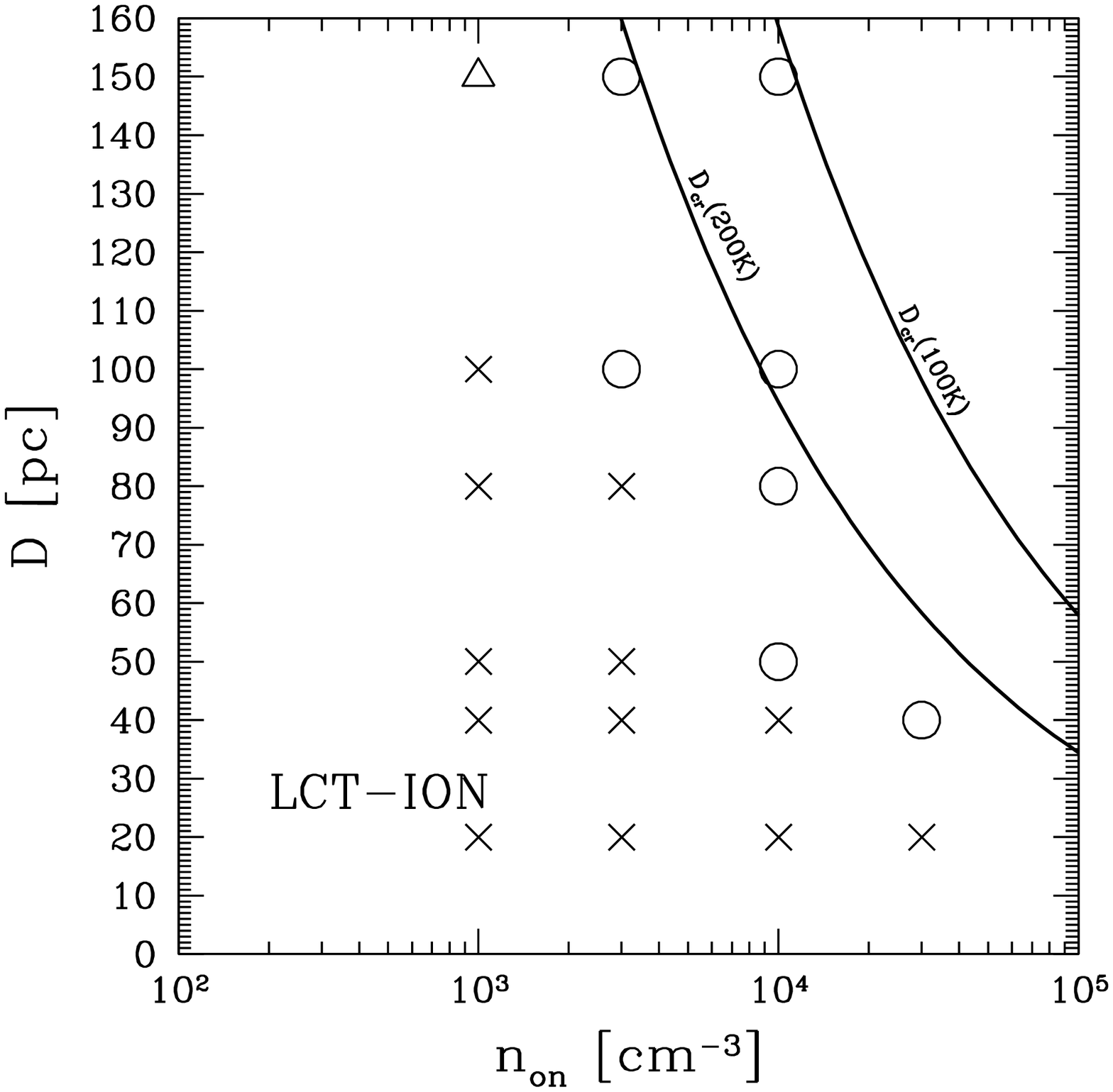}
\caption{Pop III star formation criteria in LTC models 
as a function of the central density $n_{\rm on}$ 
at turning on UV and the distance $D$ from a source star (120$M_\odot$). 
Circles succeed in collapsing, while crosses fail to
collapse. Triangles also succeed in collapsing, although they fail to
collapse in the absence of ionizing photons.
Solid curves show the collapse criterion in the presence 
of H$_2$ dissociating radiation only \citep{Susa07}. 
}\label{LTCcriterion}
\end{center}
\end{figure}

\begin{figure}
\begin{center}
\includegraphics[angle=0,width=7cm]{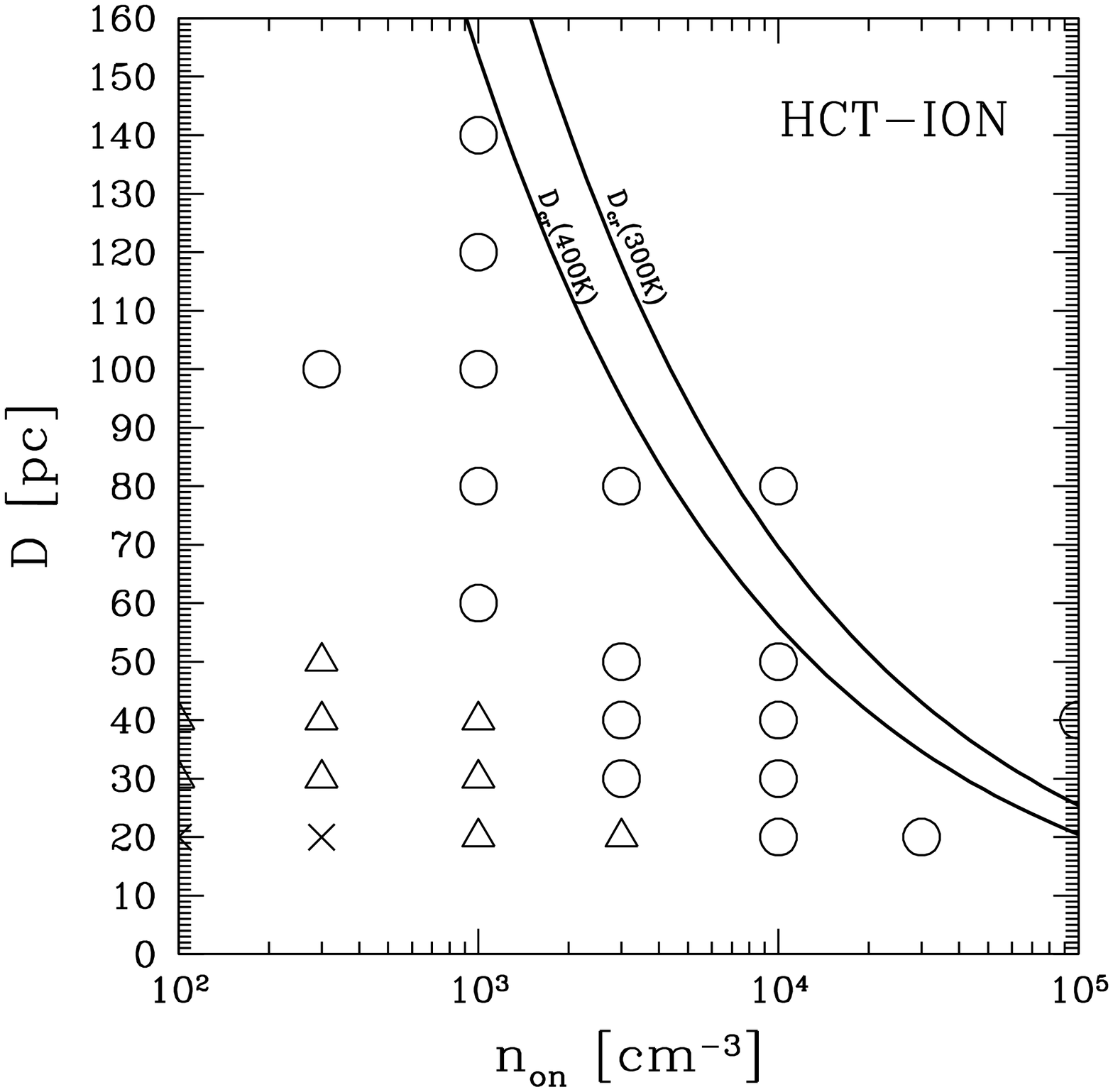}
\caption{Same as Fig.\ref{LTCcriterion}, except that the cloud
models are HTC models.
}\label{HTCcriterion}
\end{center}
\end{figure}

In addition to four runs described in the previous section, we perform
numerical simulations with various $D$ and $n_{\rm on}$. 
Figs. \ref{LTCcriterion} and \ref{HTCcriterion} show 
the results of various runs for LTC and HTC models, respectively. 
The horizontal axis represents the density $n_{\rm on}$ at turning on 
a source star and the vertical axis shows
the distance $D$ between the source star and the cloud center. 
In the runs denoted by crosses, the clouds cannot collapse until the
end of simulations, i.e. $10^7$yr after the ignition of a source star. 
The open circles represent the runs with
successful collapse. These runs can collapse even without ionizing
radiation. On the other hand, the runs denoted by open triangles can
collapse only if ionizing radiation is present.
Here, ``a collapsed cloud'' is defined as the one
whose central density exceeds the numerical resolution limit determined
by the Jeans condition. The density corresponding to the
resolution limit is $\sim 10^6 - 10^7{\rm cm^{-3}}$ in the present simulations. 
Solid curves show the collapse criterion in the presence 
of H$_2$ dissociating radiation only, which
is the same as the one obtained by \citet{Susa07}.
Figs. \ref{LTCcriterion} and \ref{HTCcriterion} give 
the threshold density depending on the distance, 
above which the cloud can keep collapsing owing to the shielding of 
H$_2$ dissociating radiation. 

Interestingly, there is no case where 
a cloud that can collapse without ionizing radiation
fails to collapse by adding ionizing radiation.
Near the threshold density, ionizing radiation can rather assist 
the collapse of a cloud that cannot 
collapse by H$_2$ dissociating radiation, as shown by open triangles
in Figs. \ref{LTCcriterion} and \ref{HTCcriterion}.
Thus, we can conclude that
ionizing flux alleviates the negative feedback by photodissociating
radiation to some extent. 
In LTC models, the parameter region of open triangles
is narrower, compared to HTC models.
The contrast basically comes from the difference in core
temperature. In both models, the H$_2$ shell shielding is almost
the same, as shown in Fig.\ref{nh2}. Also, 
the mild shock preceding the ionization front heats up the core, 
which leads to efficient H$_2$ formation and radiative cooling. 
This effect is also reported by one-dimentional RHD simulations \citep{AS07}.
However, the hydrodynamic evolution 
in the two models is very different. 
In HTC models, the central part
of the core can cool and collapse with aid of the mild shock heating as
well as the H$_2$ shell shielding. In LTC models, 
since the core mass and radius are smaller,
a diffracted shock front is converged into the core
(upper right panels in Figs.\ref{H2map}-\ref{Tmap}) before the collapse is completed.
As a result, the central part of the core is heated up again. 
This second shock heating is too strong to keep the gas cloud
bounded by self-gravity. Thus, the core bounces and evaporates 
after this second impact by a shock. 
This negative shock heating mechanism cancels the
positive effects by the H$_2$ shell shielding and mild shock heating in
LTC-ION models.

\section{Mass of secondary pop III stars}

To estimate the final
stellar mass after the radiation hydrodynamic feedback,
the mass accretion timescale is compared to
the Kelvin-Helmholtz (KH) contraction timescale, 
which is adopted from \cite{Oshea07}.

\begin{figure}
\begin{center}
\includegraphics[angle=0,width=7cm]{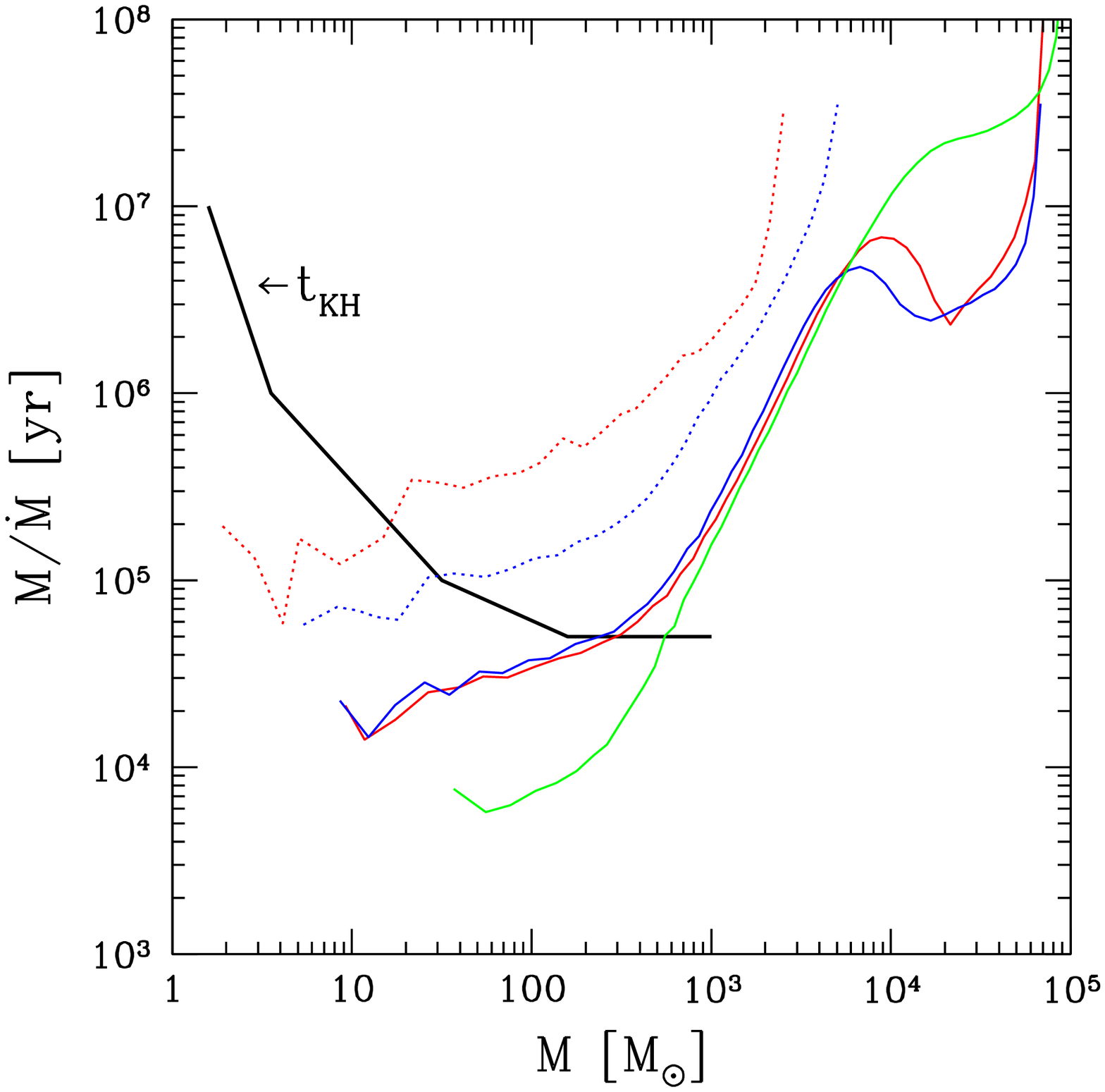}
\caption{Mass accretion timescale ($M/\dot{M}$) and 
Kelvin-Helmholtz (KH) contraction timescale
 are plotted for five models in which the cores successfully
 collapse. All models are Low Temperature Core models (i.e. $T_{\rm
 c,min}=150$K). Green line: no feedback, 
red solid line: $n_{\rm on}=10^4{\rm cm^{-3}}$ and $D=100$pc, 
red dotted line: $n_{\rm on}=10^4{\rm cm^{-3}}$ and $D=50$pc,
blue solid line: $n_{\rm on}=3\times10^3{\rm cm^{-3}}$ and $D=150$pc,
and
blue dotted: $n_{\rm on}=3\times10^3{\rm cm^{-3}}$ and $D=100$pc.
A black line is
the KH contraction timescale adopted from \cite{Oshea07}.
}
\label{KH}
\end{center}
\end{figure}

Fig. \ref{KH} shows the mass
accretion timescale as a function of enclosed mass from the center 
of the density peak for several LTC models. 
The enclosed mass is defined as the mass within a sphere of
a given radius. The mass accretion rate $\dot{M}$ is 
defined as the average of $\dot{M}$ over particles:
\begin{equation}
\dot{M}=\frac{1}{n_{\rm shell}}\sum_{i=1}^{n_{\rm shell}}4\pi\rho_i
\left(\bm{v}_i-\bm{v}_{\rm pk}\right)
\cdot(\bm{x}_{\rm pk}-\bm{x}_i)|\bm{x}_{\rm pk}-\bm{x}_i|,
\end{equation}
where $n_{\rm shell}$ denotes the number of SPH particles within a thin
shell $(r,r+\delta r)$, and the summation is taken over the particles
within the shell. $\bm{x}_{\rm pk}$ and
$\bm{v}_{\rm pk}$ denote the position and velocity of the particle
at the density peak, and $\bm{x}_i$ and $\bm{v}_i$ are 
those of the particle in the shell.

Five models in which the core collapses successfully are
plotted. Four models are LTC models with ionizing photons,  
while a model denoted by a green curve represents 
a UV-free (no feedback) case as a reference. 
In the case of no feedback, the
accretion timescale intersects the KH contraction timescale
for the enclosed mass of $\sim 6\times 10^2M_\odot$. 
This is comparable to the previous estimates 
\citep{Bromm99,NU01,Abel02,Yoshida06a}.

On the other hand, with radiative feedback, the accretion timescale
becomes longer than that in no feedback case. For instance, in the model 
with $n_{\rm on}=3\times 10^3{\rm cm^{-3}}$ and $D=100$pc (blue dotted curve), 
the mass accreted in the KH timescale is reduced to a level of 
$\sim 20M_\odot$. Therefore, the radiative feedback has a significant
impact on the mass of secondary Pop III stars. We also remark that this effect has
strong dependence on $n_{\rm on}$ and $D$. 
Blue and red dotted curves in Fig.\ref{KH}
denote the model close to the boundary of successful collapse region
(see Fig. \ref{LTCcriterion}), whereas
blue and red solid curves correspond to the models slightly distant
from the boundary. The mass accretion timescale in latter two
models is also prolonged by radiative feedback, but the effect is not
so dramatic as the former two models. 
In these cases, the expected mass
of secondary stars is as large as $\approx 100 M_\odot$.

\section{Conclusions and Discussion}

We have explored the evolution of a collapsing cloud
nearby a first luminous star to examine quantitatively 
radiation hydrodynamic feedback on secondary Pop III star formation. 
To investigate this issue, three-dimensional radiation SPH
simulations have been performed, 
where the radiative transfer of ionizing photons
and H$_2$ dissociating photons from a first star
is self-consistently incorporated. 
We have derived the numerical criteria for the subsequent 
Pop III star formation.
As a result, we have found that 
an H$_2$ shell formed ahead of an ionizing front works 
effectively to shield the H$_2$ dissociating radiation,
leading to the positive feedback. 
On the other hand, a shock associated with an ionizing front
has a positive effect and a negative effect.
The positive one is the compression and heating of gas cloud that
promote H$_2$ formation and thereby accelerate the gravitational
instability of the cloud.
The negative one is the stripping of gas envelope that 
reduces the mass accretion rate, leading to the formation
of less massive Pop III stars. 
In a small core, the compression and heating
by a shock turns to be devastating to hinder the cloud collapse.
In comparison of the mass accretion timescale with
Kelvin-Helmholtz timescale, we have found that 
the mass of secondary Pop III stars could be reduced
to $\approx 20M_\odot$.

The possibility that the formation of low-mass Pop III stars 
with $<100M_\odot$ is driven by radiation-hydrodynamic feedback  
would be important in the environment where multiple stars can form 
within a $\sim (100{\rm pc})^3$ volume. 
In the conventional one-star-per-halo picture, 
this low-mass star formation mode would not be so important. 
However, in the case of first galaxies, this mode could play 
an important role as well as the other modes driven 
by HD cooling \citep{UI00, NU02,Nagakura05,JB06,GB06,Yoshida07}. 
In addition, even in a minihalo, the assumption of one-star-per-halo itself 
could be modified by resolving dark matter evolution with very high accuracy
\citep{Umemura09}.

If such lower-mass Pop III stars are formed, the radiation-hydrodynamic feedback
on further star formation could be changed, since the importance
of ionizing radiation relative to H$_2$ dissociating radiation is different from 
that of a 120$M_\odot$ star. 
\citet{HUS09} have investigated the dependence on the stellar mass
of radiation-hydrodynamic feedback and 
concluded that the ionizing radiation from $\sim 25M_\odot$
stars does not have significant impact on the neighboring star formation,
but the feedback is essentially determined by H$_2$ dissociating radiation,
which causes only a negative effect.

In the present simulations, we omitted a few processes.
First, we do not include the HD chemistry/cooing. It is well known
that HD is an important coolant in case once the gas is heated above
$10^4$K. In that case, H$_2$ cooling proceeds faster than the
recombination of hydrogen atom at several thousand Kelvin, leading to a large H$_2$ fraction,
because of ``frozen'' electrons \citep{SK87,SUNY98,OhH02}. HD molecules are formed from these
abundant H$_2$ molecules as the collapse of cloud
proceeds\citep[e.g,][]{UI00,NU02}. Thus, HD cooling could be important 
when the ionized gas recollapses after the death of the source star.
However, in the present simulations, the collapsing density peaks 
do not experience such a hot phase (Fig.\ref{Evolution}), because
of the self-shielding of the core. The shock associated to the I-front
is also too mild to ionize the material (again Fig.\ref{Evolution}). Thus,
as far as the collapse criteria of the core is concerned,
HD cooling do not play an important role. 
Secondly, we also do not include the H$_2$-H$^+$ collision-induced H$_2$
cooling. According to \cite{GA08}, H$_2$-H$^+$ collision dominates the
excitation of H$_2$ in case electron fraction $y_{\rm e}\ga 10^{-2}$ at
$T\sim 10^4$K, and $y_{\rm e}\ga 10^{-3}$ at $T\sim 10^3$K. These
conditions are marginally satisfied when the gas cools down from ionized
state. Thus, H$_2$-H$^+$ collision-induced H$_2$ cooling can play
an important role, 
when the ionized gas recollapses after the source star turned off.
However, again in our simulations, it is not so important, since
the gas at the density peak is not ionized in its history. 
It is also worth noting that cooling and dynamics in the neighbor of H$_2$
shell could be affected by these cooling process \citep{WN08}, 
however, it is unlikely that this effect changes the present results.

\bigskip
We are grateful to all the collaborators in 
{\it Cosmological Radiative Transfer Codes Comparison Project}
for fruitful discussions during
the three workshops.
Numerical simulations have been performed with computational facilities 
at Center for Computational Sciences in University of Tsukuba
and with computational facilities in Rikkyo University. 
This work was supported in part by the {\it FIRST} project based on
Grants-in-Aid for Specially Promoted Research by 
MEXT (16002003) and Grant-in-Aid for Scientific 
Research (S) by JSPS  (20224002) and Inamori Research Foundation.



\begin{thebibliography}{}
\bibitem[Abel, Bryan, Norman(2000)]{Abel00}
Abel, T., Bryan, G. L., \& Norman, M. L. 2000, \apj, 540, 39
\bibitem[Abel, Bryan, Norman(2002)]{Abel02}
Abel, T., Bryan, G. L., \& Norman, M. L. 2002, Science, 295, 93 
\bibitem[Ahn \& Shapiro(2007)]{AS07} 
Ahn, K., \& Shapiro, P.~R.\ 2007, \mnras, 375, 881
\bibitem[Alvarez et al.(2006)]{Alvarez06} 
Alvarez, M.~A., Bromm, V., \& Shapiro, P.~R.\ 2006, \apj, 639, 621 
\bibitem[Baraffe, Heger \& Woosely (2001)]{Baraffe01}
Baraffe, I., Heger, A. \& Woosely, S.E. \  2001, \apj, 550, 890
\bibitem[Barnes \& Hut(1986)]{BH86}
Barnes, J., \& Hut, P. 1986, Nature, 324, 446 
\bibitem[Bromm, Coppi \& Larson(1999)]{Bromm99}
Bromm, V., Coppi, P. S., \& Larson, R. B. 1999, \apj, 527, L5
\bibitem[Bromm, Coppi \& Larson(2002)]{Bromm02}
Bromm, V., Coppi, P. S., \& Larson, R. B. 2002, \apj, 564, 23
\bibitem[Bromm et al.(2003)]{Bromm03} 
Bromm, V., Yoshida, N., 
\& Hernquist, L.\ 2003, \apjl, 596, L135 
\bibitem[Cen(2003)]{Cen03} Cen, R. 2003, \apj, 591, L5
\bibitem[Ciardi Ferrara, \& White(2003)]{Cia03} 
Ciardi, B., Ferrara, A., \& White, S. D. M. 2003, \mnras, 344, L7
\bibitem[Dijkstra et al.(2004)]{Dijkstra04} 
Dijkstra, M., Haiman, Z., Rees, M.~J., \& Weinberg, D.~H.\ 2004, \apj, 601, 666 
\bibitem[Draine \& Bertoldi (1996)]{DB96}
Draine, B. T., \& Bertoldi, F. 1996, \apj, 468, 269
\bibitem[Dubinski(1996)]{Dubinski96}
Dubinski, J. 1996, NewA, 1, 133 
\bibitem[Fuller \& Couchman(2000)]{Fuller00} 
Fuller, T.~M., \& Couchman, H.~M.~P.\ 2000, \apj, 544, 6 
\bibitem[Galli \& Palla (1998)]{GP98} 
Galli D. \& Palla F. 1998, \aap, 335, 403 
\bibitem[Glover \& Abel(2008)]{GA08} 
Glover, S.~C.~O., \& Abel, T.\ 2008, \mnras, 388, 1627 
\bibitem[Glover \& Brand(2001)]{GB01} 
Glover, S.~C.~O., \& Brand, P.~W.~J.~L.\ 2001, \mnras, 321, 385 
\bibitem[Greif \& Bromm(2006)]{GB06}
Greif, T.~H., \& Bromm, V., 2006, MNRAS, 373, 128 
\bibitem[Greif et al.(2007)]{Greif07} 
Greif, T.~H., Johnson, 
J.~L., Bromm, V., \& Klessen, R.~S.\ 2007, \apj, 670, 1 
\bibitem[Haiman et al.(1997)]{Haiman97} 
Haiman, Z., Rees, M.~J., \& Loeb, A.\ 1997, \apj, 476, 458 
\bibitem[Haiman et al.(2000)]{Haiman00} 
Haiman, Z., Abel, T., \& Rees, M.~J.\ 2000, \apj, 534, 11 
\bibitem[Hasegawa et al.(2009)]{HUS09} Hasegawa, K., Umemura, 
M., \& Susa, H.\ 2009, \mnras, 395, 1280 
\bibitem[Heinemann et al.(2006)]{Heinemann06} 
Heinemann, T., Dobler, W., Nordlund, A., \& Brandenburg, A. 2006, A\&A, 448, 731 
\bibitem[Iliev et al.(2006)]{Iliev06} 
Iliev, I. T., et al. 2006, MNRAS, 371, 1057 
\bibitem[Iliev et al.(2009)]{Iliev09} 
Iliev, I.~T., et al.\ 2009, arXiv:0905.2920 
\bibitem[Iwamoto et al.(2005)]{Iwamoto05} 
Iwamoto, N., Umeda, H., Tominaga, N., Nomoto, K., \& Maeda K. 2005, 
Science, 309, 451
\bibitem[Johnson \& Bromm(2006)]{JB06} 
Johnson, J.~L., \& Bromm, V., 2006, MNRAS, 366, 247 
\bibitem[Johnson et al.(2008)]{Johnson08} 
Johnson, J.~L., Greif, T.~H., \& Bromm, V.\ 2008, \mnras, 694 
\bibitem[Kahn(1954)]{Kahn54} 
Kahn, F.~D.\ 1954, \bain, 12, 187
\bibitem[Kang \& Shapiro (1992)]{KS92}
Kang, H., \& Shapiro, P., \apj, 386, 432
\bibitem[Kitayama et al.(2001)]{Kitayama01} 
Kitayama, T., Susa, H., Umemura, M., \& Ikeuchi, S.\ 2001, \mnras, 326, 1353 
\bibitem[Kitayama et al. (2004)]{Kitayama04} 
Kitayama,T., Yoshida, N., Susa, H. \& Umemura, M., 2004, \apj, 613, 631
\bibitem[Kitayama \& Yoshida(2005)]{KY05} 
Kitayama, T., \& Yoshida, N.\ 2005, \apj, 630, 675 
\bibitem[Machacek et al.(2001)]{Machacek01} 
Machacek, M.~E., Bryan, G.~L., \& Abel, T.\ 2001, \apj, 548, 509 
\bibitem[Mori et al.(2002)]{Mori02} 
Mori, M., Ferrara, A., 
\& Madau, P.\ 2002, \apj, 571, 40 
\bibitem[Murakami et al.(2005)]{Murakami05} 
Murakami, T., Yonetoku, D., Umemura, M., Matsubayashi, T., 
\& Yamazaki, R.\ 2005, \apjl, 625, L13 
\bibitem[Nakamoto, Umemura, \& Susa (2001)]{NUS01} 
Nakamoto, T., Umemura, M., \& Susa, H.\ 2001, \mnras, 321, 593 
\bibitem[Nakamura \& Umemura(1999)]{NU99} 
         Nakamura F. \& Umemura M. 1999, \apj, 515, 239
\bibitem[Nakamura \& Umemura(2001)]{NU01}
	Nakamura, F., \& Umemura, M. 2001, \apj, 548, 19
\bibitem[Nakamura \& Umemura(2002)]{NU02} 
Nakamura, F., \& Umemura, M., 2002, ApJ, 569, 549 
\bibitem[Nagakura \& Omukai(2005)]{Nagakura05} 
Nagakura, T., \& Omukai, K., 2005, MNRAS, 364, 1378 
\bibitem[Nishi \& Susa(1999)]{NS99} Nishi, R., \& Susa, H.\ 1999, \apjl, 523, L103 
\bibitem[Oh \& Haiman (2002)]{OhH02}
Oh, P \& Haiman, Z. \ 2002, \apj, 569, 558
\bibitem[Omukai \& Nishi (1999)]{ON99}
Omukai, K. \& Nishi, R. \ 1999, \apj, 518, 64
\bibitem[O'Shea \& Norman(2007)]{Oshea07} 
O'Shea, B.~W., \& Norman, M.~L.\ 2007, \apj, 654, 66 
\bibitem[Ricotti, Gnedin, \& Shull(2001)]{Ricotti01}
Ricotti, M. Gnedin, N.~Y., Shull, M. \ 2001, \apj, 560, 580
\bibitem[Ricotti \& Ostriker(2004)]{Ricotti04} 
Ricotti, M.~\& Ostriker, J.~P.\ 2004, \mnras, 350, 539
\bibitem[Scannapieco, Ferrara, \& Madau(2002)]{Scan02} 
Scannapieco, E., Ferrara, A., \& Madau, P.\ 2002, \apj, 574, 590 
\bibitem[Shapiro \& Kang (1987)]{SK87}
Shapiro, P.R., \& Kang, H., 1987, \apj, 318, 32
\bibitem[Sokasian et al.(2004)]{Sokasian04} 
Sokasian, A., Yoshida, N., Abel, T., Hernquist, L., \& Springel, V.\ 2004, \mnras, 350, 47 
\bibitem[Somerville \& Livio(2003)]{Somer03} 
Somerville, R.~S.~\& Livio, M.\ 2003, \apj, 593, 611 
\bibitem[Spitzer(1978)]{Spitzer78} 
Spitzer, L., Jr. 1978, Physical Processes in the Interstellar Medium (New York: Wiley) 
\bibitem[Steinmetz \& Muller(1993)]{SM93} 
Steinmetz, M., \& Muller, E. 1993, A\&A, 268, 391 
\bibitem[Susa (2006)]{Susa06}
Susa, H. 2006, PASJ, 58, 455 
\bibitem[Susa (2007)]{Susa07}
Susa, H.\ 2007, \apj, 659, 908 
\bibitem[Susa \& Kitayama(2000)]{SK00} 
Susa, H., \& Kitayama, T. 2000, MNRAS, 317, 175 
\bibitem[Susa et al.(1998)]{SUNY98} Susa, H., Uehara, H., 
Nishi, R., \& Yamada, M.\ 1998, Progress of Theoretical Physics, 100, 63 
\bibitem[Susa \& Umemura(2000)]{SU00} 
Susa, H. \& Umemura, M. 2000, \apj, 537, 578
\bibitem[Susa \& Umemura(2004a)]{SU04a} 
Susa, H. \& Umemura, M. 2004a, \apj, 600, 1
\bibitem[Susa \& Umemura(2004b)]{SU04b} 
Susa, H. \& Umemura, M. 2004b, \apj, 610, 5L
\bibitem[\protect\citeauthoryear{Susa \& Umemura}{2006}]{SU06} 
Susa, H., \& Umemura, M., 2006, ApJ, 645, L93
\bibitem[Tajiri \& Umemura(1998)]{TU98} 
Tajiri, Y., \& Umemura, M.\ 1998, \apj, 502, 59 
\bibitem[Tegmark et al.(1997)]{Tegmark97} 
Tegmark, M., Silk, J., 
Rees, M.~J., Blanchard, A., Abel, T., \& Palla, F.\ 1997, \apj, 474, 1 
\bibitem[Thacker et al.(2000)]{Thacker00} 
Thacker, J., Tittley, R., Pearce, R., Couchman, P., \& Thomas, A. 2000, MNRAS, 319, 619 
\bibitem[Tornatore et al.(2007)]{Tornatore07} 
Tornatore, L., Ferrara, A., \& Schneider, R.\ 2007, \mnras, 382, 945 
\bibitem[Uehara \& Inutsuka(2000)]{UI00}
Uehara, H., \& Inutsuka, S., 2000, ApJ, 531, L91
\bibitem[Umeda \& Nomoto(2003)]{Umeda03} 
Umeda, H., \& Nomoto, K., 2003, Nature, 422, 871
\bibitem[Umemura(1993)]{Umemura93} 
Umemura, M. 1993, ApJ, 406, 361 
\bibitem[Umemura et al.(2007)]{Umemura07} 
Umemura, M., Susa, H., Suwa, T., Sato, D., \& FIRST Project Team, 2007,
{\it First Stars III} (Eds. O'Shea, B.W., Heger, A., \& Abel, T.), 386
\bibitem[Umemura et al.(2009)]{Umemura09}
Umemura, M., Suwa, T., Susa, H., \& FIRST Project Team, 2009, in preparation
\bibitem[Whalen \& Norman(2008)]{WN08} Whalen, D., \& Norman, M.~L.\ 2008, \apj, 673, 664 
\bibitem[Whalen et al.(2008)]{Whalen08} 
Whalen, D., O'Shea, B.~W., Smidt, J., \& Norman, M.~L.\ 2008, \apj, 679, 925 
\bibitem[Wyithe \& Loeb(2004)]{Wyi04} 
Wyithe, J.~S.~B.,~\& Loeb, A.\ 2004, \nat, 427, 815 
\bibitem[Yoshida et al. (2003)]{Yoshida03}
Yoshida, N., Abel, T., Hernquist, L., \& Sugiyama, N., 2003, \apj, 592, 645
\bibitem[Yoshida(2006)]{Yoshida06a} 
Yoshida, N.\ 2006, New Astronomy Review, 50, 19 
\bibitem[Yoshida et al.(2006)]{Yoshida06b} 
Yoshida, N., Omukai, K., Hernquist, L., \& Abel, T.\ 2006, \apj, 652, 6 
\bibitem[Yoshida, Omukai \& Hernquist(2007)]{Yoshida07}
Yoshida, N., Omukai, K., \& Hernquist, L., 2007, ApJ, 667, L117

\end{thebibliography}
\end{document}